# Acoustic transmission enhancement through a periodically-structured stiff plate without any opening

Zhaojian He, Han Jia, Chunyin Qiu[*], Shasha Peng, Xuefei Mei, Feiyan Cai, Pai Peng, Manzhu Ke, and Zhengyou Liu

Key Laboratory of Artificial Micro- and Nano-structures of Ministry of Education and School of Physics and Technology, Wuhan University, Wuhan 430072, China

**Abstract:**

We report both experimentally and theoretically that the enhanced acoustic transmission can occur in the subwavelength region through a thin but stiff structured-plate without any opening. This exotic acoustic phenomenon is essentially distinct from the previous related studies originated from, either collectively or individually, the interaction of the incident wave with openings in previous structures. It is attributed to the structure-induced resonant excitation of the *non-leaky* Lamb modes that exist *intrinsically* in the uniform elastic plate. Our finding should have impact on ultrasonic applications.

**PACS numbers:** 43.20.+g, 42.79.Dj, 43.35.+d

---

[*] To whom all correspondence should be addressed, cyqiu@whu.edu.cn

The extraordinary optical transmission through a metallic film patterned with periodical subwavelength structure has stimulated significant interest in recent years [1-7], because of its prospective applications. The physical origin of this effect has been widely attributed to the excitation of surface plasmon [1-3] on metal surfaces, but there is still a debate on the relative contribution of the surface wave versus the Fabry-Pérot resonance [4-7]. Inspired by the studies in optical systems, investigations on acoustic waves through artificially-structured solid plates immersed in fluid have been widely reported [8-17]. Similarly, the acoustic transmission enhancement (ATE) has also been observed. For the hard plate case [8-11], the exotic phenomenon arises from the interaction of the incident wave with openings, either collectively or individually (depending on the thickness of the sample). The crossover behavior has also been observed [8-10]. A recent study states that there exists an essential difference for the contribution of the lattice resonance between the hole arrays and the slit arrays [12]. Concerning the elasticity of the perforated plates, the extraordinary shielding of sound [13] and the excitation of the intrinsic Lamb modes [14] have been reported. The possibility of the ATE through the excitation of the structure-induced acoustic surface waves in perfectly rigid body has also been discussed [15-17].

In this letter, we consider a system consisting of a water-immersed thin brass plate patterned with periodical gratings. In this stiff structured system, both the experimental and numerical results demonstrate that the ATE can occur in the subwavelength region. We have verified that the exotic phenomenon arises from the resonant excitation of the *non-leaky* Lamb modes that exist *intrinsically* in uniform solid plates, a close analog to the transmission enhancement associated with the surface plasmon excitation in its optical counterpart. In particular, in the present system there does not exist any opening, and hence it is different from those structures used previously with the opening as a precondition to obtain the ATE.

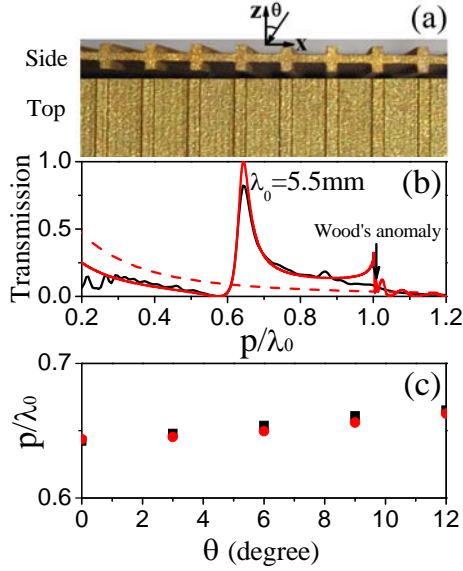

FIG. 1 (color online). (a) Side and top views of a segment of the sample fabricated by patterning periodical rectangular gratings on both sides of a uniform brass plate. (b) The experimental (dark solid line) and numerical (red solid line) power transmissions at normal incidence plotted as a function of the normalized frequency $p/\lambda_0$ for the structured sample, accompanied with the one for a uniform reference plate (red dashed line), where $p$ and $\lambda_0$ are the structural period and the wavelength in water, respectively. (c) The experimental (dark squares) and numerical (red circles) frequencies of the resonant peaks for different incident angles $\theta$.

As shown by the images in Fig. 1(a), the sample consists of a thin brass plate (thickness $t = 0.7mm$) patterned with a periodical array (period $p = 3.5mm$) of the rectangular brass gratings (width $0.9mm$ and height $0.7mm$) on both sides. The total width and length of the sample are $8cm$ and $14cm$ (thereby covering 40 periodical units), respectively. In the experiment, the sample is clamped with desired tilted angles, and placed in between a couple of ultrasonic generating and detecting transducers (with central frequencies of $0.5MHz$ and diameters of $2.5cm$). The detecting transducer is aligned normally towards the generating one with a distance of $30cm$. The entire assembly is immersed in a big water tank. The power transmission is obtained through a well-known ultrasonic transmission technique [9,13,14]. The experimental data presented below are obtained by averaging over three independent measurements. Throughout this letter, we use the finite-difference time-domain method [18] to numerically calculate the transmission spectrum, dispersion relation, and field distribution for the structured sample.

In Fig. 1(b), we present the experimental (dark solid line) and numerical (red solid line) transmission spectra at normal incidence for the structured sample. Both of them show a remarkable peak with frequency well below the Wood's anomaly (indicated by the arrow) that corresponds to $p = \lambda_0$, with $\lambda_0$ being the wavelength in water. This is in contrast to the corresponding transmission spectrum (red dashed line) for the uniform reference plate of thickness $t = 0.7mm$, where the transmission is rather low in the frequency range under consideration because of the large acoustic impedance contrast between brass and water. The experimental data agrees excellently with the theoretical one as a whole, especially for the position of the transmission peak. The discrepancy of the transmission amplitude could mostly stem from the imperfections of the structure and the limited precision in the experimental measurement. We have also measured and calculated the transmission spectra at several small incident angles. The transmission peak is robust, and its position shows a blue shift as the angle increases, as shown in Fig. 1(c).

Note that the wavelength of the transmission peak ($\lambda_0 \simeq 5.5mm$) is obviously larger than any characteristic lengths of the sample, such as the depth ($0.7mm$) of the rectangular grooves, the thickness of the brass plate ($0.7mm$), and the period of the brass gratings ($3.5mm$). Therefore, the peak is neither due to the Fabry-Pérot resonances of the grooves or the flat plate itself, nor due to the coherent diffraction of the periodical gratings. In fact, the diffraction peak near the Wood's anomaly is very sharp and not observed in the experiment [see Fig. 1(b)]. Intuitively, the ATE could stem from the resonant excitation of some kind of intrinsic non-leaky modes in the uniform plate.

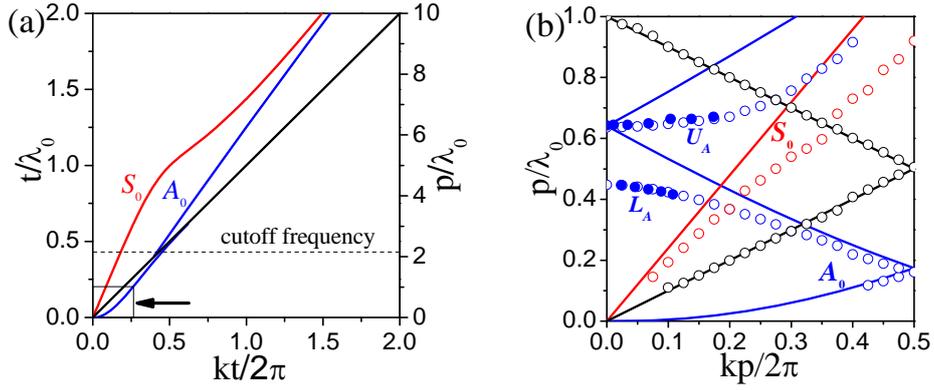

FIG. 2 (color online). (a) Dispersion curves of the $S_0$ (red line) and $A_0$ (blue line) modes for a uniform brass plate of thickness $t$ immersed in water, accompanied with the water line (dark line). Here the horizontal dashed line denotes a cutoff frequency below which the $A_0$ mode is non-leaky. (b) Numerical dispersion curves (open circles) for the structured sample, together with the ones (solid circles) extracted from the (numerical) transmission peaks at several incident angles. For comparison, the simply-folded dispersion curves [within the rectangular framework in (a)] for the uniform plate are replotted in (b) with the same color lines.

Intrinsically, a free solid plate supports Lamb modes [14,19] with the vibration energy confined inside it. When the plate is immersed in a fluid background, the case changes a little bit: the energies of the Lamb modes turn out to be leaky or not, mostly depending on their phase velocities (in contrast to that of fluid). The Lamb modes can be classified into two categories by their symmetries of the substance motion with respect to the midplane of the plate: the symmetric modes $S_n$ and the antisymmetric modes $A_n$, with $n$ (integer) characterizing their orders. Here we focus on the two zero-order ones, i.e., $S_0$ and $A_0$ modes, since only they exist all the way down to the zero frequency, while the higher-order modes exhibit lower-cutoff frequencies [14,19]. Both the $S_0$ and $A_0$ modes are strongly dispersive due to the finite plate thickness $t$, as shown in Fig. 2(a). For the $S_0$ mode, the dispersion curve lies always above the water line and hence it is leaky. For the $A_0$ mode, however, below a cutoff frequency the dispersion curve lies below the water line: it is non-leaky and can't be excited by the external plane wave incident at any direction due to the mismatch of momentum. Physically, the low phase velocity of the non-leaky $A_0$ mode (NLAM) is a consequence of the stiffness-thickness relationship for a thin plate in bending: the plate gets effectively softened as its thickness reduces.

In order to find the relationship between the transmission peaks and the above mentioned intrinsic modes, the dispersion curves of the structured sample have been numerically simulated by the finite-difference time-domain method, which are displayed by open circles in Fig. 2(b). Compared with the simply-folded ones for the uniform plate (lines), although the shapes are considerably deformed, the dispersion curves denoted by the red and blue open circles stem convincingly from the $S_0$ and $A_0$ modes, respectively [20]. Because of the Bragg scattering by the gratings, the dispersion curve of the $A_0$ mode divides into a couple of dispersion branches $U_A$ and $L_A$, separated by a band gap. By matching the transversal component of the wavevector, in Fig. 2(b) we also present the peak positions (blue solid circles) extracted from the numerical transmission spectra for several incident angles [in good agreement with the experimental ones, see Fig. 1(c)]. We find that the transmission peak corresponds exactly to the resonant excitation of the mode on the upper dispersion branch $U_A$. As indicated by the transmission spectra in Fig. 1(b), the mode on the lower dispersion branch $L_A$ is not excited at normal incidence because of the mismatch of the symmetries between the external plane wave and the eigen-mode [21]. Even at oblique incidence, the modes on $L_A$ are weakly excited such that the corresponding transmission peaks are extremely sharp and can only be observed in the numerical transmission spectra, which are also picked out and presented in Fig. 2(b).

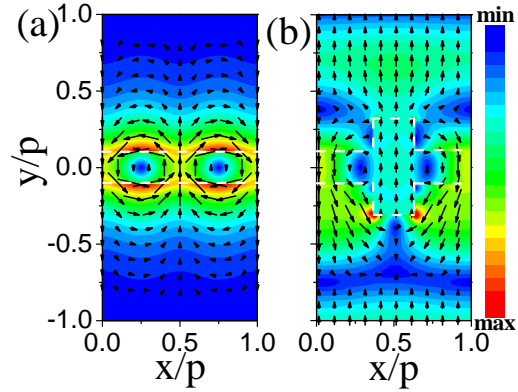

FIG. 3 (color online). (a) A typical distribution of the displacement field for the $A_0$ mode in a uniform plate. (b) The displacement field distribution for the sample excited by a normally incident plane wave at the resonant frequency. Both the amplitude field (color variations) and the vector field (arrows) are provided. Here the white dashed lines indicate the brass-water interface in a periodical unit.

Now an interesting question arises: why the ATE originated from the resonant excitation of the NLAM has not been observed (or announced) in similar works, even in the case of the structured *brass* plate immersed in *water* [9,11,13]. We find that the presence/absence of openings in the structure plays an important role.

Figure 3(a) displays a typical displacement field distribution for the NLAM in a uniform plate. It is observed that the vibration field inside the plate forms acoustic vortices, which clearly demonstrates the importance of the shear component in the solid. In this work, the solid bumps act as good couplers between the external wave and the intrinsic NLAM [22] because of the full use of the shear character of the NLAM, as shown in Fig. 3(b), the displacement field distribution excited by a plane wave normally incident upon the sample at the resonant frequency. However, in the slit/hole structures reported previously, the incident (longitudinal) waves can transmit through the openings directly without strong interactions with the NLAMs. Our further numerical investigations for such structures state that the transmission peak related with the NLAM is very sharp and difficult to be observed in the experiment.

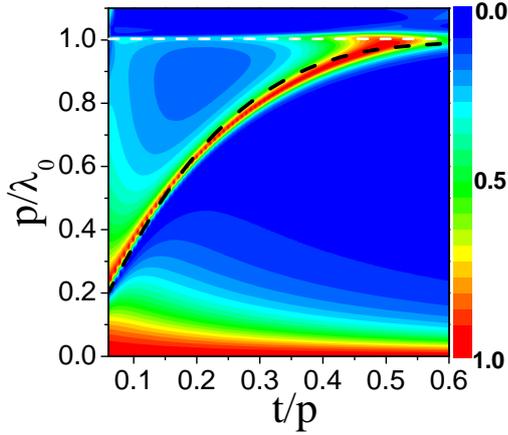

FIG. 4 (color online). Transmission spectra of normal incidence plotted as a function of $t/p$ for a series of the samples with gratings patterned identically to the aforementioned. Here the dark dashed line gives a theoretical prediction of the resonant frequency from the non-leaky $A_0$ mode in the uniform brass plate of thickness $t$, obtained by applying a fictitious periodicity $p$ for the momentum transformation.

Because of the strongly dispersive character of the NLAM in the low frequency region [see Fig. 2(a)], the normalized frequency of the transmission peak greatly depends on the ratio of the plate thickness $t$ to the structural period $p$, which even goes to the deep subwavelength region if the plate is thin enough. This can be seen in Fig. 4, the transmission spectra varying with the ratio $t/p$ for a series of the samples with gratings patterned identically to the aforementioned. The position of the transmission peak agrees well with the one predicted from the NLAM in the uniform plate, obtained by applying a fictitious periodicity for the momentum transformation, as shown by the dark dashed line. It increases monotonously with $t/p$ because of the effective stiffening of the solid plate, and approaches to a constant of $p/\lambda_0 = 1$ since at the cutoff frequency the phase velocity of the NLAM is identical to the water velocity [see Fig. 2(a)], merging together with the peak induced by lattice resonance nearby the Wood's anomaly. An inverse tendency has been found in the structures with openings, where the ATE arises from the diffraction-assisted Fabry-Pérot resonance [8-10]. [That conclusion can also be made in a hard solid limit since the vibration concentrates mostly in the fluid, greatly different from our case where the vibration in solid takes a dominant role.] The resonant excitations of the leaky Lamb modes occur at higher frequencies ($p/\lambda_0 > 1$) for a large value of $t/p$ [14], which are out of the scope of this work and not displayed here.

In conclusion, with using a new type of structures, we have observed the ATE by exciting the non-leaky Lamb modes in uniform plates. It is worth noting that the exotic phenomenon can occur in deep subwavelength region, which is different from the case with regard to the leaky Lamb modes [14]. Prospective applications of this new mechanism to achieve the ATE can be anticipated, such as ultrasonic filters and ultrasonic medical instrumentations.

**Acknowledgement:**

This work is supported by the National Natural Science Foundation of China (Grant Nos. 10874131, 10974147, and 10944003); and NSFC/RGC joint research grants 10731160613 and N_HKUST632/07.

dispersion curves for the sample approach to the ones of the uniform plate when the scattering gratings are tuned smaller and smaller.

[21] I. R. Hooper and J. R. Sambles, Phys. Rev. B **70**, 045421 (2004); S. G. Rodrigo, L. Martin-Moreno, A. Y. Nikitin, A. V. Kats, I. S. Spevak, and F.J. Garcia-Vidal, Opt. Lett. **34**, 4 (2009).

[22] Of course, the coupling strength depends on the size/shape of the solid bumps. In particular, the double-side corrugation considerably benefits the coupling between the incident wave and the NLAM, compared with single-side corrugated samples (see the supplementary material).